\newcommand{\metaAuthorMail}{alexander.willner@fokus.fraunhofer.de}
\newcommand{\metaAuthorShort}{A. Willner}

\newcommand{\metaSubject}{A Structured Description of the Distributed Cloud Computing Paradigm for Manufacturing Use Cases}
\newcommand{\metaTitle}{Towards a Reference Architecture Model for Industrial Edge Computing}
\newcommand{\metaKeywords}{Distributed Computing, Cloud Computing, Edge Computing, Fog Computing, IoT, IIoT, Industry 4.0, Smart Manufacturing}

\RequirePackage{fix-cm}

\documentclass[12pt,technote]{IEEEtran}
\usepackage[backend=biber,
  style=numeric-comp,%
  maxnames=3,
  maxcitenames=3,
  maxbibnames=3,
  doi=true,
  isbn=false,
  url=false,
  ]{biblatex}
\usepackage{amsmath,amssymb,amsfonts}
\usepackage{algorithmic}
\usepackage{graphicx}
\usepackage{textcomp}
\usepackage{xcolor}
\usepackage{mathptmx}

\usepackage{url}
\usepackage{acro}
\newcommand\acc[1]{%
{%
\renewcommand{\url}[1]{}%
\renewcommand{\footnote}[1]{}%
\renewcommand{\cite}[1]{}%
\renewcommand{\nobreakspace}{}%
\let\protect\relax%
\ac*{#1}%
}%
}
\newcommand\aclc[1]{%
{%
\renewcommand{\url}[1]{}%
\renewcommand{\footnote}[1]{}%
\renewcommand{\cite}[1]{}%
\renewcommand{\nobreakspace}{}%
\let\protect\relax%
\acl*{#1}%
}%
}

\definecolor{LinkColor}{rgb}          %
{0.31,0.46,.64}                       %
\definecolor{MarginColor}{rgb}        %
{0.31,0.46,.64}                       %

\usepackage{ifoddpage}                %
\usepackage{needspace}                %
\usepackage{marginnote,zref-savepos}  %

\newcommand\sidenote[1]{\mbox{}%
  \marginnote{%
    \scriptsize%
    \hspace{0pt}%
    \color{MarginColor}%
      \emph{#1}%
  }%
}%
\makeatletter
\@ifpackageloaded{tex4ht}{\renewcommand\sidenote[1]{}}{}
\makeatother

\usepackage{blindtext}                %

\newif\ifacrousefootnote%
\DeclareAcronym{CPS}{short={CPS}, long={Cyber-physical System}}
\DeclareAcronym{5G}{short={5G}, long={5th Generation Mobile Network}, cite={ngmn-5g}}
\DeclareAcronym{IoT}{short={IoT}, long={Internet of Things}}
\DeclareAcronym{RAMEC}{short={RAMEC}, long={Reference Architecture Model Edge Computing}}
\DeclareAcronym{RAMI}{short={RAMI4.0}, long={Reference Architecture Model Industrie 4.0}, cite={din91345}}
\DeclareAcronym{ICT}{short={ICT}, long={Information and Communication Technology}}
\DeclareAcronym{IIoT}{short={IIoT}, long={Industrial Internet of Things}}
\DeclareAcronym{SCL}{short={SCL}, long={Structured Control Language}}
\DeclareAcronym{IEC}{short={IEC}, long={International Electrotechnical Commission}}
\DeclareAcronym{MES}{short={MES}, long={Manufacturing Execution System}}
\DeclareAcronym{I/O}{short={I/O}, long={Input/Output}}
\DeclareAcronym{SCADA}{short={SCADA}, long={Supervisory Control and Data Acquisition}}
\DeclareAcronym{ERP}{short={ERP}, long={Enterprise Resource Planning}}
\DeclareAcronym{SPS}{short={SPS}, long={Speicherprogrammierbare Steuerung}, long-plural-form={Speicherprogrammierbaren Steuerungen}}
\DeclareAcronym{AAS}{short={AAS}, long={Asset Administration Shell}, short-indefinite={an}, long-indefinite={an}}
\DeclareAcronym{KI}{short={KI}, long={Künstliche Intelligenz}}
\DeclareAcronym{HMI}{short={HMI}, long={Human Machine Interface}}
\DeclareAcronym{AIOTI}{short={AIOTI}, long={Alliance for Internet of Things Innovation}}
\DeclareAcronym{ETSI}{short={ETSI}, long={European Telecommunications Standards Institute}}
\DeclareAcronym{KMU}{short={KMU}, long={kleine und mittlere Unternehmen}}
\DeclareAcronym{IIC}{short={IIC}, long={Industrial Internet Consortium}, long-post={\footnote{\url{http://iiconsortium.org}}}}
\DeclareAcronym{IDS}{short={IDS}, long={Intrusion Detection System}}
\DeclareAcronym{CDN}{short={CDN}, long={Content Delivery Network}}
\DeclareAcronym{RAMEC4.0de}{short={RAMEC4.0}, long={Referenz Architektur Modell Edge Computing 4.0}}
\DeclareAcronym{SGAM}{short={SGAM}, long={Smart Grid Architecture Model}, cite={sgam2012}}
\DeclareAcronym{RAMIde}{short={RAMI}, long={Referenzarchitekturmodell Industrie 4.0}, cite={pi4rami2015,din91345}}
\DeclareAcronym{TSN}{short={TSN}, long={Time-Sensitive Networking}, long-post={\footnote{\url{http://ieee802.org/1/pages/tsn.html}}}}
\DeclareAcronym{VLC}{short={VLC}, long={Visible Light Communication}}
\DeclareAcronym{LoRaWAN}{short={LoRaWAN}, long={Long Range Wide Area Network}}
\DeclareAcronym{NB-IoT}{short={NB-IoT}, long={Narrowband IoT}}
\DeclareAcronym{HART}{short={HART}, long={Highway Addressable Remote Transducer}}
\DeclareAcronym{AMQP}{short={AMQP}, long={Advanced Message Queuing Protocol}}
\DeclareAcronym{OPC}{short={OPC}, long={Object Linking and Embedding for Process Control}}
\DeclareAcronym{WoT}{short={WoT}, long={Web of Things}}
\DeclareAcronym{PKI}{short={PKI}, long={Public Key Infrastructure}}
\DeclareAcronym{RTOS}{short={RTOS}, long={Real-Time Operating System}}
\DeclareAcronym{TPU}{short={TPU}, long={Tensor Processing Unit}}
\DeclareAcronym{TOE}{short={TOE}, long={TCP Offload Engine}}
\DeclareAcronym{VM}{short={VM}, long={Virtual Machine}}
\DeclareAcronym{EECC}{short={EECC}, long={European Edge Computing Consortium}, long-post={\footnote{\url{https://ecconsortium.org}}}}
\DeclareAcronym{OT}{short={OT}, long={Operational Technology}, long-plural-form={Operational Technologies}}
\DeclareAcronym{Cloud Computing}{short={Cloud Computing}, long={Cloud Computing}, cite={mell2011nist}, class=exclude}
\DeclareAcronym{Edge Computing}{short={Edge Computing}, long={Edge Computing}, cite={edgecomputing2015,edge2016}, class=exclude}
\DeclareAcronym{ICS}{short={ICS}, long={Industrial Control System}}
\DeclareAcronym{PLC2}{short={PLC}, long={Programmable Logic Controller}}
\DeclareAcronym{STL}{short={STL}, long={Statement List}}
\DeclareAcronym{LD2}{short={LD}, long={Ladder Diagram}}
\DeclareAcronym{PAC}{short={PAC}, long={Programmable Automation Controller}}
\DeclareAcronym{IPC}{short={IPC}, long={Industrial PC}}
\DeclareAcronym{OPC UA}{short={OPC UA}, long={Open Platform Communications Unified Architecture}}
\DeclareAcronym{vPLC}{short={vPLC}, long={Virtual Programmable Logic Controller}}
\DeclareAcronym{AR2}{short={AR}, long={Augmented Reality}}
\DeclareAcronym{DIN}{short={DIN}, long={Deutsche Industrie Norm}}
\DeclareAcronym{MEC}{short={MEC}, long={Multi-Access Edge Computing}}
\DeclareAcronym{IEEE}{short={IEEE}, long={Institute of Electrical and Electronics Engineers}}
\DeclareAcronym{LPWAN}{short={LPWAN}, long={Low-Power Wide-Area Network}}
\DeclareAcronym{HTTP}{short={HTTP}, long={Hyper Text Transfer Protocol}}
\DeclareAcronym{CoAP}{short={CoAP}, long={Constrained Application Protocol}}
\DeclareAcronym{MQTT}{short={MQTT}, long={Message Queue Telemetry Transport}}
\DeclareAcronym{DDS}{short={DDS}, long={Data Distribution Service}}
\DeclareAcronym{VPN}{short={VPN}, long={Virtual Private Network}}
\DeclareAcronym{TPM}{short={TPM}, long={Trusted Platform Module}}
\DeclareAcronym{GPU}{short={GPU}, long={Graphics Processing Unit}}
\DeclareAcronym{FPGA}{short={FPGA}, long={Field Programmable Gate Array}}
\DeclareAcronym{VLAN}{short={VLAN}, long={Virtual LAN}}
\DeclareAcronym{LXC}{short={LXC}, long={LinuX Container}}
\DeclareAcronym{SME}{short={SME}, long={Small and Medium Enterprise}, short-indefinite={an}}
\DeclareAcronym{AI}{short={AI}, long={Artificial Intelligence}}
\DeclareAcronym{MANO}{short={MANO}, long={Management and Orchestration}}
\DeclareAcronym{IIRA}{short={IIRA}, long={Industrial Internet Reference Architecture}, cite={iira}}
\DeclareAcronym{Grid Computing}{short={Grid Computing}, long={Grid Computing}, class=exclude}
\DeclareAcronym{SDO}{short={SDO}, long={Standards Developing Organization}}
\DeclareAcronym{PTCC}{short={PTCC}, long={PubSub TSN Centralized Configuration}}
\DeclareAcronym{CNC}{short={CNC}, long={Central Network Controller}}
\DeclareAcronym{CUC}{short={CUC}, long={Centralized User Configuration}}
\DeclareAcronym{VNF}{short={VNF}, long={Virtualized Network Function}}
\DeclareAcronym{SDN}{short={SDN}, long={Software-Defined Networking}}
\DeclareAcronym{CPPS}{short={CPPS}, long={Cyber-physical Production System}}
\DeclareAcronym{DINen}{short={DIN}, long={German Industrial Standard}}
\DeclareAcronym{ROI}{short={ROI}, long={Return on Investment}}
\DeclareAcronym{AGV}{short={AGV}, long={Autonomous Guided Vehicle}}
\DeclareAcronym{SCI4.0}{short={SCI4.0}, long={Standardization Council Industrie 4.0}}
\DeclareAcronym{OMG}{short={OMG}, long={Object Management Group}}
\DeclareAcronym{MSP}{short={MSP}, long={Multi Stakeholder Platform}}
\DeclareAcronym{DER}{short={DER}, long={Distributed Energy Ressource}}
\DeclareAcronym{3GPP}{short={3GPP}, long={3rd Generation Partnership Project}}
\DeclareAcronym{URLLC}{short={URLLC}, long={Ultra-Reliable Low-Latency Communication}}
\DeclareAcronym{WAN}{short={WAN}, long={Wide Area Network}, short-indefinite={a}, long-indefinite={a}}
\DeclareAcronym{GMPLS}{short={GMPLS}, long={Generalized Multi Protocol Label Switching}}

\makeatletter
\@ifpackageloaded{tex4ht}{\usepackage[tex4ht]{hyperref}}{
\usepackage[%
]{hyperref}                           %
}
\makeatother

\hypersetup{
      pdfauthor={\metaAuthorShort~<\metaAuthorMail>},
      pdfsubject={\metaSubject},
      pdftitle={\metaTitle},
      pdfkeywords={\metaKeywords},
      pdfcreator={http://github.com/tubav/paper},
      pdfproducer={pdfTeX},
}

\makeatletter
\renewcommand{\@seccntformat}[1]{}
\makeatother

\usepackage{tikz,stackengine,mathtools,changes}
\usetikzlibrary{matrix,calc}
\newsavebox{\foobox}

\definechangesauthor[name={Alexander Willner}, color=orange]{AW}
\definechangesauthor[name={Varun Gowtham}, color=blue]{VG}

\usepackage[pscoord]{eso-pic}
\newcommand{\ieeecopyright}[3]{
 \setbox0=\hbox{#3}
 \AddToShipoutPictureFG*{ \put(\LenToUnit{#1\paperwidth},\LenToUnit{#2\paperheight}){\vtop{{\null}\makebox[0pt][c]{#3}}}}
}
\ieeecopyright{.2}{0.055}{978-1-5386-5541-2/18/\$31.00~\copyright~2020~IEEE}

\begin{document}

\title{\metaTitle
}

\author{Alexander Willner\thanks{Alexander Willner is with Fraunhofer FOKUS, Berlin, Germany} and Varun Gowtham\thanks{Varun Gowtham is with Technische Universität Berlin, Berlin, Germany}
}

\maketitle

\begin{abstract}
  In the context of the digital transformation of the industry,
  whole value chains get connected across various application domains;
  as long as economic, ecologic, or social benefits arise to do so.
Under the umbrella of the \acc{IIoT},  
  traditional \acc{OT} approaches are replaced or at least augmented by
  \acc{ICT} systems to facilitate this development.
To meet industrial requirements,
  for example, related to privacy, determinism, latency, or autonomy,
  established \aclc{Cloud Computing} mechanisms are being moved closer to data sources and actuators.
Depending on the context,
  this distributed \aclc{Cloud Computing} paradigm is named \aclc{Edge Computing} or Fog Computing
  and various challenges have been subject to several publications.
However,
  a proper reference model 
  that describes the multi-dimensional problem space which is being spanned by this paradigm,
  seems still to be undefined.
Such a model should provide orientation,
  put work in relation and 
  support the identification of current and future research issues.
This paper aims to fill this gap
  with a focus on industrial automation
  and follows analog models that have been developed for specific domains 
  such as the \acc{SGAM} and the \acc{RAMI}.
The proposed \acc{RAMEC} identifies 210 views on the Edge Computing paradigm 
  in the manufacturing domain.
Future iterations of this model 
  might be used for the classification of relevant 
  research, standardization, and development activities.
\acresetall
\end{abstract}


\section{Introduction}\label{sec:intro}

\sidenote{Context: IIoT}
To optimize the efficiency,
  to establish new business models or
  to enhance sustainability,
  various value chains are in the process of getting digitally connected
  using \ac{IoT} technologies.
Within industrial domains,
  the established \ac{OT} is in the process of being augmented or replaced by \ac{ICT}.
This evolution is sometimes denoted as the \ac{IIoT}.
One driving factor is the motivation to operate a converged, more software-based communication infrastructure,
  built upon open standards,
  to enhance flexibility
  and to allow for data-driven business models
  in the production industries.
This does not only reduce costs for operators
  but also facilitates a shorter time-to-market span for new products.

\sidenote{Challenge: Requirements}
Specific challenges in this context 
  are requirements related to industrial communication systems.
To support determinism, low latency, and autonomy,
  many technological approaches related to field buses 
  and \acp{ICS} have been evolved in the last 70 years.
A significant milestone was the invention of the \ac{PLC2} in 1968
  to easily modify local control loops.

\sidenote{Problem: Limitations}
However,
  programming \acp{PLC2} did not evolve at the same pace
  as Internet technologies.
Many control systems are still programmed using \acp{STL}, \acp{LD2}, or a \ac{SCL},
  following the \ac{IEC} 61131 standards.
Not only do these approaches impose significant limitations
  to the potential computing capabilities modern endpoints expose,
  but also younger generations face significant challenges updating existing code.
Further,
  the maintainability, scalability, and modularity of these programs are rudimentary.

\sidenote{Existing Work: IEC 61499, PACs}
To address the latter,
  the \ac{IEC} 61499 standards introduce an event-driven distributed system
  based on the \ac{IEC} 61131.
Further,
  \acp{PAC}
  enable the use of higher-level instructions in parallel to 
  software-based \acp{PLC2} for hard real-time control, for example on \acp{IPC}.

\sidenote{Approach: Edge Computing}
These developments are now being augmented 
  by recent developments from within the \ac{ICT} domain,
  more specifically from the fields of network and system management
  and distributed systems and communication networks respectively.
In the 70 years history of computing,
  we can observe a cyclic alternation
    between a centralized computing paradigm (1950s: Mainframes; 2000s: Cloud Computing)
  and a distributed computing paradigm (1980s: Client Server).
The current Edge Computing paradigm~\cite{LopezEdge2005} distributes the Cloud Computing paradigm %
  by moving functionalities again closer to data sources and actuators.
This allows for the usage of Cloud Computing technologies 
  within several critical \ac{IoT} domains, %
  particularly in industrial use cases.
For example,
  complex applications or federated learning algorithms~\cite{kairouz2019advances}
  can be orchestrated towards Edge nodes
  to either influence local control loops or
  to provide valuable data to a \ac{MES}.

\sidenote{Contribution: RAMEC}
However,
  a review of related work indicated
  that the multi-dimensional problem space,
  that the usage of the Edge Computing paradigm in industrial fields of application spans,
  has not been properly described yet.
Therefore,
  the main contribution of this paper 
  is the presentation of a \acf{RAMEC} for the manufacturing domain.
Analog to the \ac{SGAM} and the \ac{RAMI}
  and in contrast to technical architectures such as \ac{MEC} by the \ac{ETSI},
  the \ac{RAMEC} is supposed to provide orientation and to put activities 
  such as initiatives, standards, publications and implementations 
  into a broader context.

\section{Related Work}\label{sec:related}

\subsection{Industrial Automation}\label{sec:pyramid}

\sidenote{Automation Pyramid}
To put the work at hand into context,
  it is important to understand the application domain we are considering,
  as different use cases entitle distinct requirements.
In this paper,
  we focus on \ac{IIoT} domains,
  i.e. the application of \ac{IoT} technologies in industrial fields of application.
More specifically,
  the automation within discrete manufacturing.
The most important concept here
  is the so-called Automation Pyramid (IEC 62264) that defines the different layers
  which modern industrial automation is composed of.

\begin{figure}[htbp]
  \centering
  \includegraphics[width=.45\textwidth]{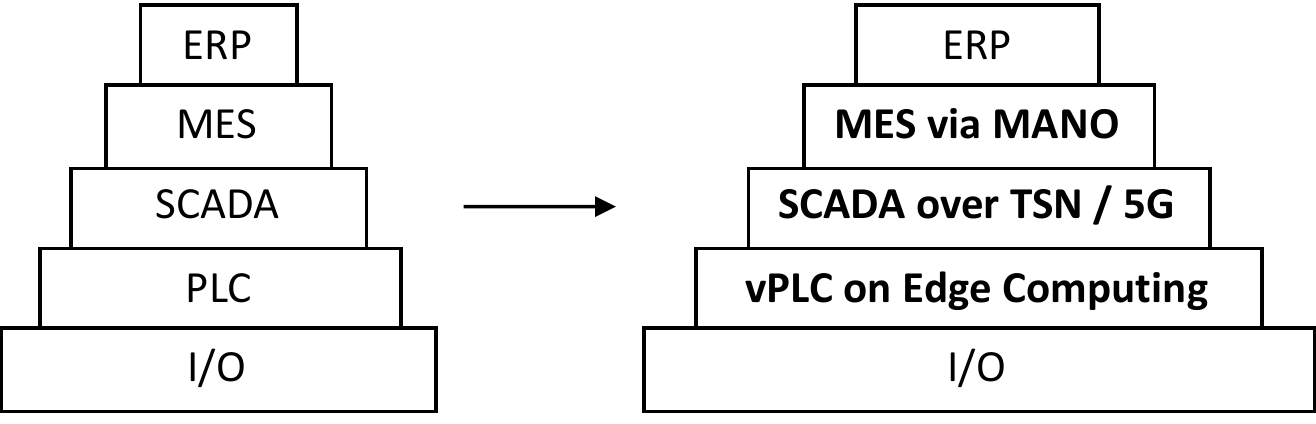}
  \caption{Future technologies within the current Automation Pyramid layers}\label{fig:pyramid}
\end{figure}

\sidenote{Figure~\ref{fig:pyramid}}
Although this pyramid is sometimes differently defined, %
  in its core, it consists of five layers (see the left part of Figure~\ref{fig:pyramid}).
It starts with an \ac{I/O} layer
  that is controlled by \acp{PLC2}.
These \acp{PLC2} are interconnected via a \ac{SCADA} system
  with the \ac{MES} layer.
Finally,
  the overall production and business systems
  are exchanging information within an \ac{ERP} system.
It is also important to notice
  that different layers are concerned with different sizes and time frames of data.

\sidenote{Industry 4.0}
Following the Industry 4.0 vision, 
  this pyramid will be replaced in the long term by the use of \acp{CPS} or \acp{CPPS}.
Intelligent units, such as \acp{AGV}, will then interact directly with each other 
  to enable the most flexible production at the lowest costs possible.
In this context, 
  the use of an \ac{AAS} is often mentioned, 
  which, in combination with an ''asset'', 
  then represents an intelligent Industry 4.0 component.

\sidenote{Towards software-based infrastructures}
To facilitate this transition,
  new software-based industrial infrastructure technologies
  are needed to be introduced into the manufacturing domain (see the right part of Figure~\ref{fig:pyramid}).
The boundaries between physical \ac{PLC2} and the classical \textit{\ac{IPC}}
  will further be blurred by the use of edge-based and distributed \acp{vPLC}. For example, based on \ac{IEC} 61499.
To meet requirements related to reliable, low-latency, and deterministic wired communication,
  the open standards within the \ac{IEEE} \ac{TSN} group are currently being adopted.
For wireless communication,
  the \ac{TSN} standards are in the process of being included in the next \ac{3GPP} release
  to allow for \ac{URLLC} over 5G networks.
Further,
  instead of only operating a centralized \ac{MES},
  a \ac{MANO} system might dynamically install and run applications on Edge nodes close to the data source,
  and, for example, \ac{AI} based pre-processing and data analytics can be executed locally~\cite{iec2017edgeintelligence}.
A handy example is an \ac{AR2} based \ac{HMI},
  which requires very short communication latencies and places high processing demands on object recognition.

\subsection{Reference Architecture Model Industry 4.0}\label{sec:rami}

\begin{figure}[htbp]
  \centering
  \includegraphics[width=.48\textwidth]{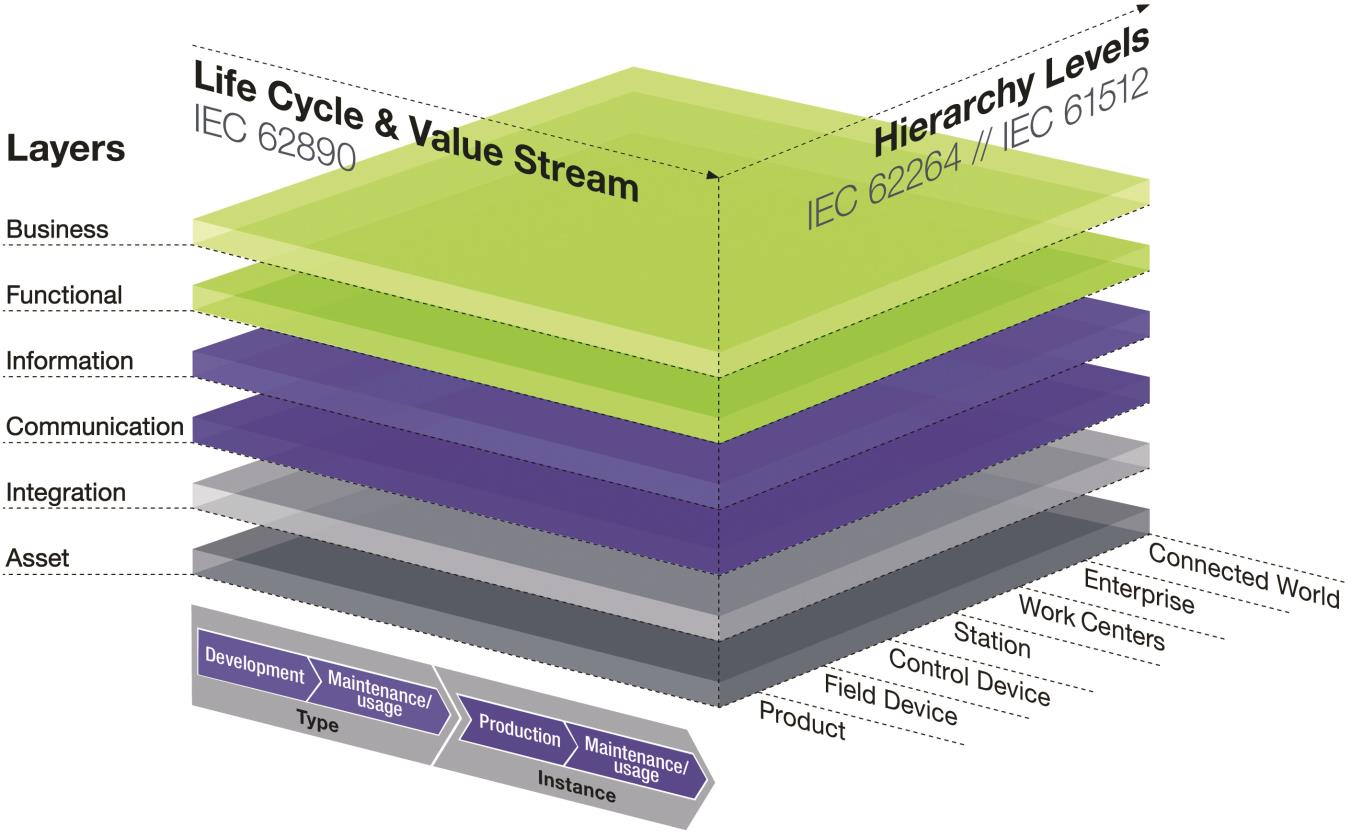}
  \caption{Reference Architecture Model Industry 4.0 (RAMI4.0)~\cite{din91345}}\label{fig:rami}
\end{figure}

\sidenote{Figure~\ref{fig:rami}}
To provide orientation and to guide technical discussions in this rather broad context,
  a reference architecture model is needed.
Within the German initiative "Plattform Industrie 4.0"
  the previously mentioned automation pyramid 
  has not only been extended in 2015 by considering the product and connected world as hierarchy levels following \ac{IEC} 62264 / 61512
  (z-axis: Product, Field Device, Control Device, Station, Work Centers, Enterprise, Connected World),
  but also they are put into relationships with various layers (y-axis: Asset, Integration, Communication, Information, Functional, Business)
  and life cycle streams following \ac{IEC} 62890 (x-axis: Development and Maintenance/usage Type, Production and Maintenance/usage Instance).
As a result, 
  the \ac{RAMI} model has been published which is depicted in Figure~\ref{fig:rami}.
It has also been standardized as \ac{DINen} SPEC 91345
  and its different aspects have been mapped~\cite{Lin2018} to the \ac{IIRA} developed by the \ac{IIC}.

\subsection{Smart Grid Architecture Model}\label{sec:sgam}
\begin{figure}[htbp]
  \centering
  \includegraphics[width=.48\textwidth]{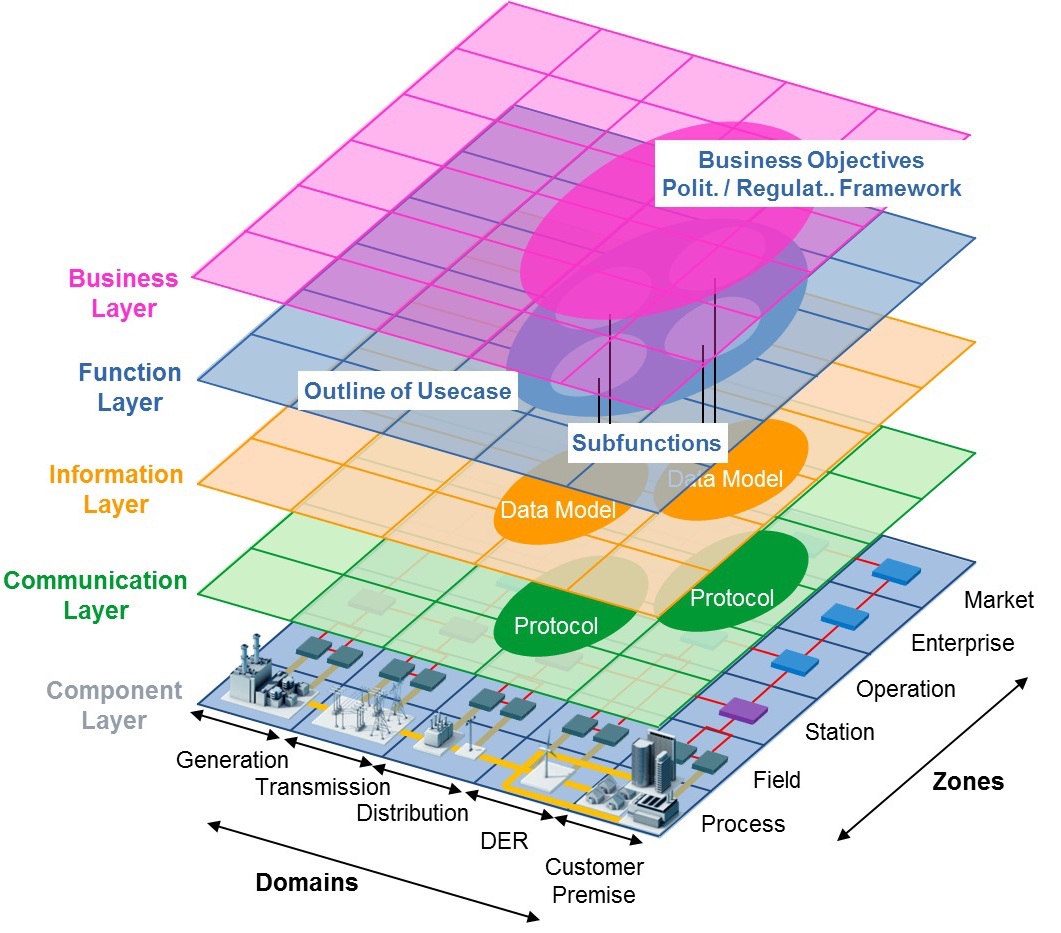}
  \caption{\acf{SGAM}}\label{fig:sgam}
\end{figure}

\sidenote{Figure~\ref{fig:sgam}}
The general idea behind the \ac{RAMI}, however,
  has been adopted previously.
Namely the \ac{SGAM} model that has been published in 2012
  to describe the important aspects relevant to smart energy networks.
As shown in Figure~\ref{fig:sgam} various zones (Process, Field, Station, Operation, Enterprise, Market),
  domains (Generation, Transmission, Distribution, \ac{DER}, Customer Premise)
  and interoperability layers (Component, Communication, Information, Function, Business) are put into relationships.
Within these intersections, protocols, data models and use cases
  are classified for a better definition of key areas.

\subsection{3D IoT Layered Architecture}\label{sec:aioti}
\begin{figure}[htbp]
  \centering
  \includegraphics[width=.5\textwidth]{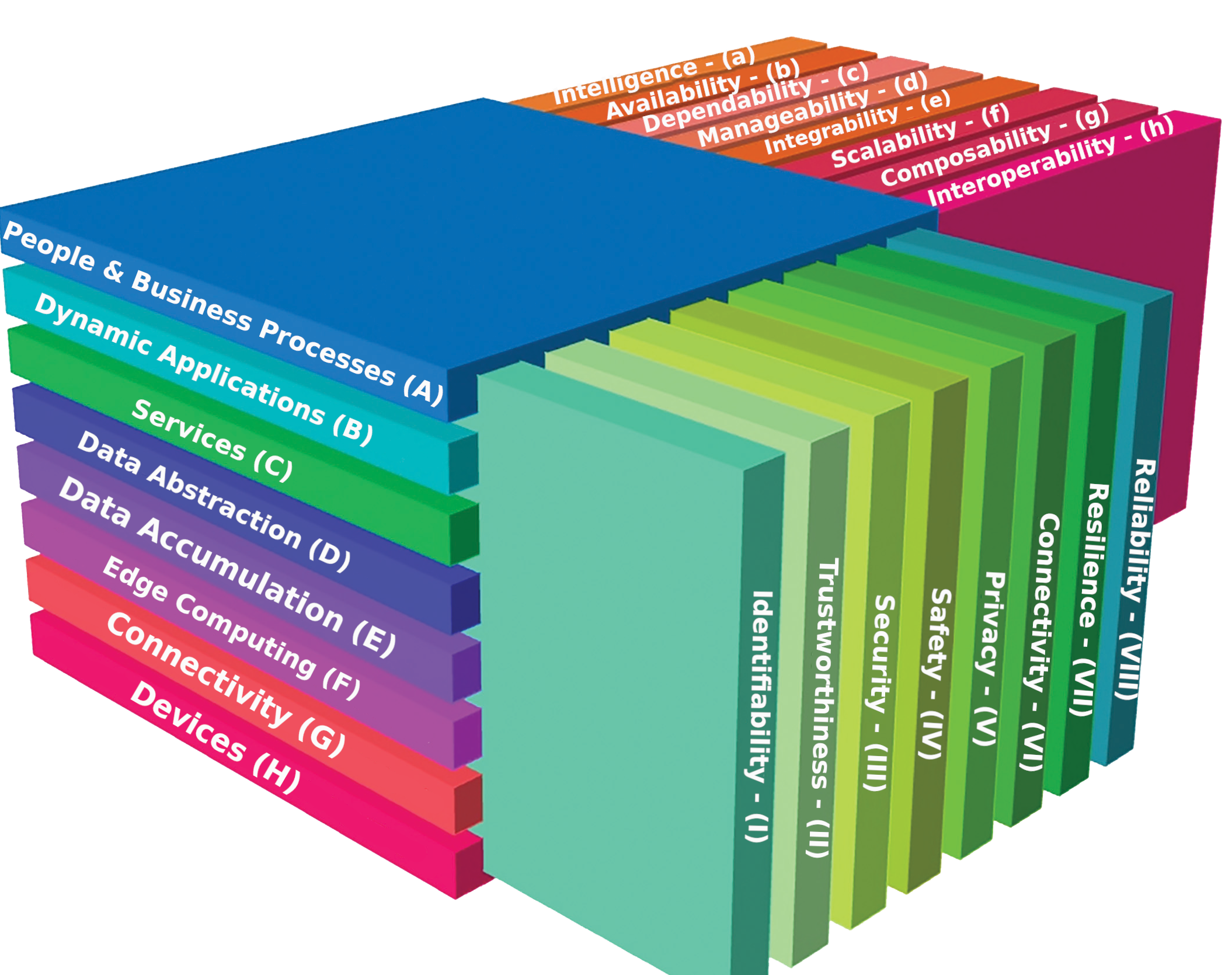}
  \caption{3D IoT Layered Architecture~\cite{vermesan2018next} (with enlarged labels)}\label{fig:aioti}
\end{figure}
\sidenote{Figure~\ref{fig:aioti}}
Based on this idea
  the \ac{AIOTI} has also widened the scope 
  and defined a 3-dimensional \ac{IoT} layered architecture in 2018.
As shown in Figure~\ref{fig:aioti},
  this 8x8x8 matrix aims at covering different concepts
  that interact with \ac{IoT} use cases.
For readability,
  we have enlarged the axes labels. They can be further identified by their respective alphabetical designation:
  on the y-axis (bottom to top): Devices (H), Connectivity (G), Edge Computing (F), Storage (E), Abstraction (D), Service (C), Applications (B), People and Processes Layers (A);
  on the x-axis (left to right): Identifiability (I), Trustworthiness (II), Security (III), Safety (IV), Privacy (V), Connectivity (VI), Resilience (VII), Reliability (VII);
  and on the z-axis (left to right): Intelligence (a), Availability (b), Dependability (c), Manageability (d), Integrability (e), Scalability (f), Composability (g), Interoperability (h).
The third layer within this figure is titled "Edge Computing"
  which is responsible in analyzing and transforming data elements.
As this distributed Cloud Computing paradigm is at the core of the paper at hand,
  it is worth to shortly review related work concerning this concept.

\subsection{Edge Computing}\label{sec:edge}

\sidenote{Definition}
While,
  generally speaking,
  Edge Computing is a distributed architecture 
  that moves storage and computation capabilities closer to data sources and actuators,
  surprisingly, the term itself has so far eluded a uniform definition.
Further,
  depending on the context and temporal embedding of the source,
  other terms such as Fog Computing, Mist Computing, or Cloudlets are being used 
  and defined in quite diverse ways as well.
In \cite{bellavista2019differentiated} the authors provide further details
  about the differentiation of these terms.

\sidenote{History}
Peter Levine, in his presentation in 2017, had embedded Edge Computing fairly well, in the bigger context of Distributed Computing.
For about 70 years we observe an alternating trend in distributed computer systems.
While the first mainframes could already be used centrally in the 1950s and 60s, 
  the trend changed in the 1980s and 90s in favor of distributed client-server systems.
One prominent example in this context is the development of \acp{CDN},
  in which data is stored and accessed in a distributed manner 
  and closer to where it is needed.
After initial research under the umbrella of the distributed \acl{Grid Computing} paradigm,
  data and services have again been stored and hosted more and more centrally 
  under the term Cloud Computing, since the beginning of the new millennium.
As mentioned before, however, 
  not all data can and should be processed outside of the own administrative domain 
  and outside defined geopolitical boundaries.
The reasons for processing data locally or centrally are manifold
  and mainly relate to the "V's of Big Data": 
  Volume (growth of data size),
  Variety (various types of data),
  and Velocity (increase in speed in which the data must be processed).
However, note that, depending on the source,
  the number of "V's" might differ between three~\cite{zikopoulos2011understanding}
  and nine~\cite{owais2016extract}, with various gradations
  (e.g. Veracity, Validity, Variability, Volatility, Visualization, and Value).
In the context of industrial automation, however, further requirements
  such as latency and determinism are added as well.

\sidenote{Literature}
Within the literature,
  one of the first occurrences of the term Edge Computing can be found in 2001,
  where the authors of~\cite{microedge2001} briefly described future challenges
  for microprocessor design.
In one of the most cited papers,
  the authors of \cite{LopezEdge2005} highlighted the future challenges in Edge Computing in 2005.
Since then,
  numerous surveys have been published %
  and the rising number of related papers indicates high research interest.

\sidenote{Standardization}
Additionally,
  \acp{SDO} have started to address this issue as well.
For example,
  within \ac{ETSI} the \ac{MEC} working group has published over 35 specifications since 2015, with~\cite{esti-mec202wlan} being the latest one,
  and within the \ac{IEC} the topic of Edge Intelligence was discussed in~\cite{iec2017edgeintelligence}.
A broader overview,
  with a focus on sustainability, 
  is given by the authors in~\cite{willner2019edgesustainability}.
In this paper, an earlier version of the \ac{RAMEC} model has briefly been described for the first time,
  however, not to the extent of this publication.

\sidenote{Adopted Definition}
Within this publication,
  we follow the Edge Computing definition
  that has been published by the Linux Foundation in August 2019.
This definition has widely been adopted within the relevant community,
  as various organizations and projects have been involved in writing 
  the corresponding glossary: %
\begin{quote}
``The delivery of computing capabilities to the logical extremes of a network in order to improve the performance, operating cost and reliability of applications and services. By shortening the distance between devices and the cloud resources that serve them, and also reducing network hops, edge computing mitigates the latency and bandwidth constraints of today's Internet, ushering in new classes of applications. In practical terms, this means distributing new resources and software stacks along the path between today's centralized data centers and the increasingly large number of devices in the field, concentrated, in particular, but not exclusively, in close proximity to the last mile network, on both the infrastructure and device sides.''~\cite{lfedge2019}
\end{quote}

\sidenote{Implementations}
Finally,
  it is important to note that the Edge Computing paradigm
  has not only raised interest in the research and standardization context.
Due to its potential business impact,
  several commercial and non-commercial implementations have been developed.
Without any claim to completeness,
  the following examples should provide an initial overview
  and are subject to change.

\sidenote{Commercial Platforms}
Commercial platforms,
  that actively support and promote the Edge Computing paradigm,
  among others are SAP Leonardo Edge Computing, Siemens MindSphere Industrial Edge,
  Microsoft Azure IoT Edge, Amazon Web Services IoT Greengrass, and IBM Watson IoT Platform Edge.
At the same time, multiple open source solutions already exist,
  that can be used, to implement the first projects.

\sidenote{Foundations}
For example, 
  under the roof of the Linux Foundation %
  individual projects are bundled together.
These include Akraino Edge Stack, Baetyl, EdgeXFoundry, 
  Edge Virtualization Engine, Fledge, Home Edge, and the before mentioned
  Open Glossary of Edge Computing.
Analogously, 
  the project FogFlow has been established under the roof of the FIWARE Foundation, %
  and the projects ioFog, Fog05, and BaSyx under the umbrella of the Eclipse Foundation. %
Within the OpenStack Foundation %
  requirements and use cases are analyzed and a reference architecture is defined
  within the Edge Computing Group.
Further,
  in the KubeEdge project, the Kubernetes framework is being extended accordingly.

\sidenote{Existing Infrastructures}
Further examples with a focus on the use of existing (mobile network) infrastructures include
  MobiledgeX, OpenEdge, Edge Gravity, or the CDNetworks Edge Computing Platform.
But also processor-related activities such as 
  LEDGE from the arm-related Linaro project are active in this field.
One of the goals is, among others, 
  to achieve extremely low reaction latency (in the range of microseconds).
Distributed edge data centers are acting as a backbone for computing capabilities 
  of a local 5G network 
  with \acp{CPS} having only limited computation power.

\section{Reference Architecture Model Edge Computing}\label{sec:ramec}

\subsection{Hierarchy Levels}\label{sec:ramec:hierarchy}

\sidenote{Hierarchy}
The brief overview above gives a first indication
  of the potential problem space to describe the Edge Computing paradigm;
  i.e. the complexity of the subject at hand, the Edge Computing paradigm,
  is rather broad and diverse.
One reason is that different terminology is being used
  and individual requirements are formulated
  in various application domains.
Another reason lies in the unspecific definition of the previously mentioned phrase
  ``the logical extremes of a network''.
Depending on the use case, the requirements, and the business model,
  the logical extreme of a network may vary widely.

\begin{figure*}
  \centering
  \includegraphics[width=.8\textwidth]{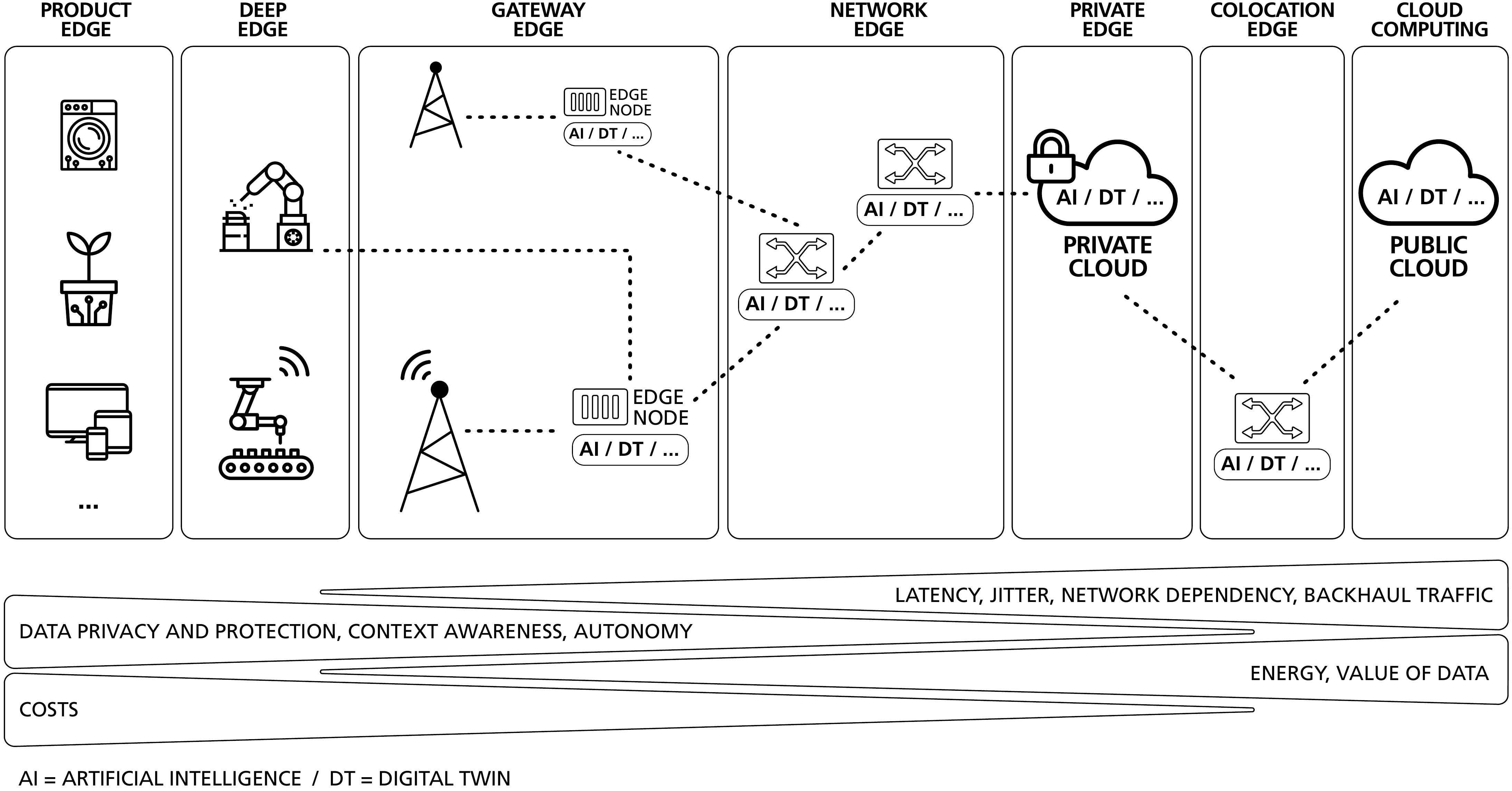}
  \caption{Edge Computing Hierarchy Levels}\label{fig:ramec:hierarchy}
\end{figure*}

\sidenote{Figure~\ref{fig:ramec:hierarchy}}
The data processing does not necessarily have to take place
  at the topologically outermost extreme.
Rather, there is a broad continuum of possible positioning in a network
  (often denoted as Fog Computing).
In Figure~\ref{fig:ramec:hierarchy}
  this spectrum of feasible distributions is depicted.
As outlined in the lower part of the figure,
  the location of Edge Computing functionality
  follows the relevant requirements and does not constitute a binary decision.
Following the illustration, at least seven different positions can be distinguished,
  which are briefly explained below.

\sidenote{Product Edge}
The \textit{Product Edge} is located directly on an intelligent product,
  which was produced for further use, e.g. a smartphone.
Another interesting application is the field of Smart Packaging,
  in which, for example, the GPS coordinates are regularly transmitted
  for track and trace use cases.
The intelligent product,
  which controls the production process individually when entering the factory,
  is more visionary.

\sidenote{Deep Edge}
The term \textit{Deep Edge} refers to the integration of cloud computing concepts
  into, for example, the production plants themselves.
Together with the intelligent product
  such as intelligent and autonomous units can interact directly with each other
  and allow for the most flexible production system possible;
  up to an individualized lot size of 1 at the cost of a mass product.

\sidenote{Gateway Edge}
The focus of many discussions and most use cases,
  however, relate to the concept of \textit{Gateway Edges}.
One reason is that
  expensive industrial assets have to be used for a long period
  to optimize the \ac{ROI}.
From an economic point of view,
  often machines are not replaced by more modern variants for several decades.
In this case,
  a gateway, that communicates with the devices by wire or wireless,
  is located nearby the asset; 
  analog to a \ac{PLC2}.

\sidenote{Network Edge}
The \textit{Network Edge} goes one step further,
  in which every node in a network is considered a potential Edge node.
For example, each switch can provide sufficient resources,
  to run smaller applications.
Examples include services  %
  such as an \ac{IDS} or a \ac{MQTT} broker.

\sidenote{Private Edge}
The \textit{Private Edge} can be put on a level with a Private Cloud.
Here the topological edge is defined as the border of the administrative domain,
  where the own area of influence ends.
This is currently within the focus of most of the commercial offerings.

\sidenote{Co-location Edge}
A particularly interesting use case is the \textit{Co-location Edge},
  sometimes referred to as Cloudlets.
Smaller data centers are arranged topologically in such a way,
  that certain requirements can be complied with.
For this,
  either existing distributed infrastructures are being used (\acp{CDN}, mobile network infrastructures, ...)
  or investments have to be made into the construction of new infrastructure.
In such a setup, both,
  the economies of scale that apply to Cloud Computing
  and the technical feasibility for deterministic communication can be combined
  (e.g. via dedicated paths using \ac{GMPLS} or wide-area \ac{TSN}).
This allows for offerings,
  such as real-time control loops in the cloud(let),
  that would not be impossible otherwise.

\subsection{Problem Space}\label{sec:ramec:space}

\sidenote{Missing Orientation}
The positioning of Edge functionalities alone shows,
  that the Edge Computing paradigm covers a wide field.
Even initiatives, articles and surveys often fail to define,
  what kind of Edge Computing is subject to the discussions.
Depending on the classification, very different requirements are placed on
  software and hardware.
Conversely, this means that,
  while various existing deployment scenarios could be located in one or multiple positions in such a continuum,
  different use cases pose even more complex requirements and not a single solution can be used in all contexts.
This leads to the main observation
  that orientation is needed to facilitate discussions.

\begin{figure}
  \centering
  \includegraphics[width=.45\textwidth]{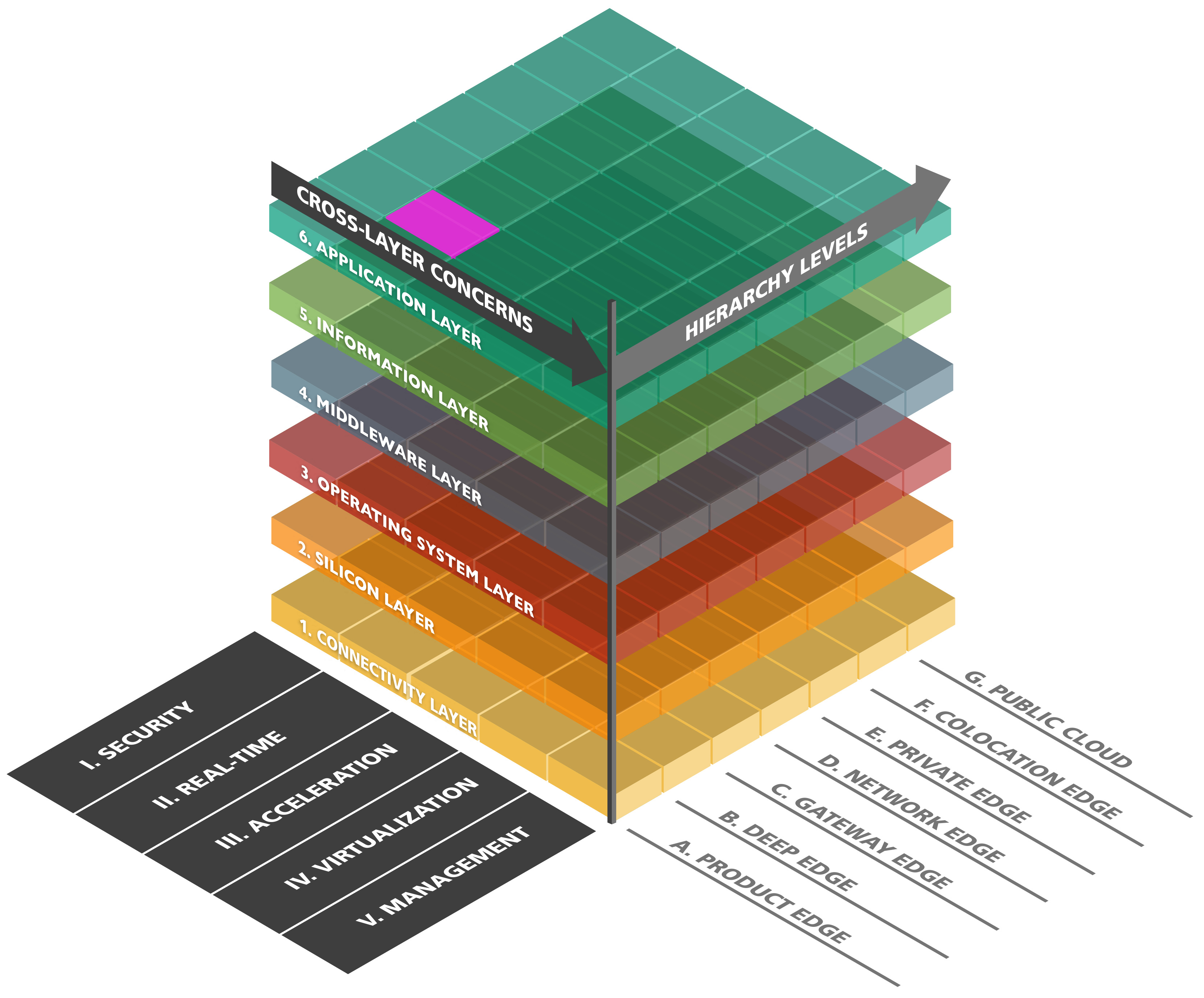}
  \caption{\acf{RAMEC}}\label{fig:ramec}
\end{figure}

\sidenote{More Dimensions}
Further,
  as indicated in the section before, %
  multiple technology layers (such as connectivity, middleware, or applications)
  have to be considered in the Edge Computing context.
Additionally,
  cross-layer concerns related to use case specific requirements
  (such as real-time behavior, security, or management)
  often influence the focus of discussions.
Therefore, 
  analog to the reference models 
  \ac{RAMI}, \ac{SGAM}, and the \ac{AIOTI} 3D \ac{IoT} Layered Architecture,
  we introduce the \acf{RAMEC}.
As shown in Figure~\ref{fig:ramec},
   it spans a 3-dimensional matrix with
   5 concerns, 6 layers, and 7 levels.
Overall,
  this provides $210$ different topics of interest within the Edge Computing paradigm.
As an example,
  the highlighted area \texttt{B. II. 6.}
  relates to real-time applications embedded within a production asset.
Another example related to \texttt{C. III. 2./6.},
  might be a \ac{TPU} accelerated edge node nearby a production line, 
  that enables quality assurance of the process based on local video stream analytics.
Further information that describes the relevant dimensions in more detail follows.

\subsection{Layers}\label{sec:ramec:layers}

\sidenote{Connectivity}
Technologies,
  that allow the Edge node to communicate with the outside world 
  (either internally to devices or externally),
  are located at the \textit{connectivity} layer.
In the industrial context, this includes developments from the 50s (4-20mA current interfaces),
  the 80s (field bus systems such as PROFIBUS, CANopen or Sercos I/II)
  and the 2000s (industrial Ethernet systems like PROFINET, EtherCat or EtherNet/IP);
  as well as but also local sensor-actuator communication technologies like IO-Link.
In the wired area, however, standards of the \ac{IEEE} from the \ac{TSN} working group are gradually being adopted.
At the same time, more and more wireless technologies are moving into the industry.
There is currently great interest in so-called 5G campus networks in particular - 
  i.e. private 5G networks in licensed frequency bands.
Also, there are suggestions to use \ac{VLC} to connect Edge nodes,
  and for longer distances fiber optic technologies
  or \acp{LPWAN} like \ac{LoRaWAN} (unlicensed band) 
  or \ac{NB-IoT} (licensed band).

\sidenote{Silicon and OS}
Traditionally, x86-based systems running Microsoft Windows have been widely used in the industry,
  that have neither been connected to the Internet nor actively updated (every minute of downtime in production costs).
With the advent of open standards and broader hardware requirements,
  it can be observed, that the underlying \textit{Silicon} architecture becomes more heterogeneous
  (e.g. the use of ARM, SPARC, or RISC-V based systems)
  and also Linux derivatives are used as \textit{Operating System}.

\sidenote{Middleware}
Many current developments can be assigned to the \textit{Middleware} layer.
While power interfaces and field bus systems require their own communication stacks (e.g. the \ac{HART} protocol),
  modern approaches mainly use the TCP/IP or UDP/IP stack of the operating system.
Examples for industrial Edge node middleware include the \ac{OPC UA},
  as well as the \ac{DDS}, oneM2M, or \ac{WoT}.

\sidenote{Information}
Ultimately, the most important resource is the actual data,
  which is assigned to the \textit{Information} layer,
  and, depending on the context, may represent so-called digital twins.
While field buses bring their own data models,
  in \ac{OPC UA} an object-oriented data model is used as a basis
  and under the term ''Companion Specifications'', currently \ac{OPC UA} related semantics are defined,
  to make it easier for machines to communicate with each other.
Other approaches like oneM2M or \ac{WoT} adapt serialization and middleware independent standards,
  designed in the Semantic Web context over the last 30 years,
  and enable the definition of distributed data structures based on linked data concepts.

\sidenote{Application}
To extract a value from this data,
  the actual applications are executed in the \textit{Application} layer.
As mentioned at the beginning, this can be 
  any kind of application or network function,
  for example, software-based \acp{PLC2} or \ac{AI} applications;
  the latter enables intelligent decision making at the edge.
This application can interact locally and autonomously with the environment
  or even synchronize with Cloud systems.
Embedded in a suitable ecosystem, new business models can emerge,
  similar to current 'AppStore' developments.

\subsection{Cross-Layer Concerns}\label{sec:ramec:crosslayers}

\sidenote{Intro}
Furthermore, requirements 
  such as security, real-time capability, (\ac{AI}) acceleration, and management
  can be addressed differently at different levels,
  which can have different characteristics depending on the topological position 
  (so-called \textit{Cross-Layer Concerns}).

\sidenote{Security}
The \textit{Security} layer includes a variety of technologies,
  to ensure the safety of an Edge node.
Depending on the level, 
  this includes Virtual Private Networks (\acs{VPN}), \acp{TPM}, Unikernels, SELinux, a \ac{PKI}, 
  and many other concepts, including blockchains.
In particular, in the context of trustworthy Edge Computing,
  a combination of isolation and slicing approaches 
  with cryptographic measures is needed to allow for distributed secure data processing.

\sidenote{Real-Time}
As mentioned before, there are often hard \textit{Real-Time} requirements, in particular, in the discrete manufacturing context.
These again have to be addressed differently on all levels and include technologies like 5G, 
\ac{TSN} or Real-Time Operating Systems (\acs{RTOS}) or the use of dedicated microcontrollers.

\sidenote{Acceleration}
For the efficient use of neural networks, for example,
  Edge nodes can be extended in the \textit{Acceleration} layer.
This may involve the use of generic Graphics Processing Units (\acs{GPU}) 
  or Field Programmable Gate Arrays (\acs{FPGA}),
  but also the use of a \ac{TPU} or other commercial \ac{AI} accelerators.
But also, classical extensions like a \ac{TOE} 
  would be located at this cross-layer concern.

\sidenote{Virtualization}
The \textit{Virtualization} layer takes into account,
  that the Edge Computing paradigm is a distributed Cloud Computing paradigm.
Hardware is shared with multiple applications,
  which are themselves executed separately again.
That means approaches like \ac{VLAN}, VT-x, \ac{LXC}, classical \acp{VM}, 
  or sandboxing play an important role here.

\sidenote{Management}
Finally, Edge nodes are distributed and must be controlled at the \textit{Management} level.
This includes the configuration of network paths and Network Slices, 
  updating firmware or the operating system, on-boarding,
  and the orchestration of applications (e.g. \acp{VNF}).
The integration into an IoT platform also takes place at this level.
A specific example is a \ac{CUC} system to configure a \ac{TSN}-based network via a \ac{CNC},
  a specific instance of to the \ac{SDN} paradigm,
  and the Edge nodes via \acs{OPC UA} using the \ac{PTCC} protocol,
  which is currently being developed within the \ac{IEEE} \acs{OPC UA} PubSub \ac{TSN} Working Group.

\subsection{Example}\label{sec:ramec:example}
\sidenote{Steel Manufacturing}
Not all of these $210$ different topics of interest within the Edge Computing paradigm
  are relevant in every use case at the same time.
To provide some context,
  there are already real-world application scenarios in the industry.
For example,
  till only a few years ago, 
  steel manufacturers developed new products through extensive tests in their research laboratories.
Today,
  this can be done faster and more efficiently as powerful computer systems take over a large part of the calculations.
To efficiently analyze the high volume of generated data,
  necessary IT capacities are sometimes installed directly on the production sites with the aid of Edge data centers.
This additional computing power enables short latency times in data provision as well as uninterrupted data availability and system-wide security.
To comply with various security requirements,
  potential steel containers that hold the IT systems
  can be equipped with security doors and have detailed monitoring for many relevant parameters,
  including access control, fire protection and fail-safe operation.

\section{Conclusions and Future Work}\label{sec:outlook}

\sidenote{Edge Computing in Manufacturing}
We have discussed the need for the Edge Computing paradigm
  in an industrial context,
  mainly focusing on manufacturing and automation.
As the interest in this area rises,
  the number of related publications, projects, and initiatives continuously grow as well.

\sidenote{Need for Orientation}
At the same time,
  more and more aspects are covered 
  which have to be taken into consideration
  when discussing the Edge Computing subject.
Unfortunately,
  a reference model,
  describing the overall industrial Edge Computing problem space,
  is undefined as of yet.
Such a reference model would support the classification of current research, 
  could put related work into relation,
  and might help to identify potential open research and development questions.

\sidenote{RAMEC}
Analog to existing reference models that already describe other problem space domains,
  such as \ac{RAMI}, \ac{SGAM}, or the \ac{AIOTI} 3D \ac{IoT} Layered Architecture,
  we have introduced the \acf{RAMEC} to fill this gap for the industrial Edge Computing context.
We gave a brief overview of its current state
  and provided several examples for each dimension.
As a result,
  $210$ distinct interdependent fields have been identified.

\sidenote{Outlook}
While this model has been discussed and matured since 2018
  with many industry partners and initiatives,
  its specific characteristics are subject to change slightly.
For instance,
  the introduction of a trusted substrate layer is currently under discussion.
  for enhanced trustworthiness reflection.  
Finally,
  as the key success factor for the acceptance of any technology in the industry,
  is standardization and easy integration into the existing infrastructure,
  the standardization of the model is planned.
For this upcoming work of the international standardization,
  several \acp{SDO} have already been contacted to identify potential next steps.
  Namely,
    this includes \ac{ETSI}, \ac{IEEE}, \ac{IEC}, the \ac{SCI4.0}, \ac{DINen}, the \ac{OMG}, and the \ac{MSP}.

\section*{Authors}
\sidenote{Alexander Willner}
Dr.-Ing. Alexander Willner [M’08] (alexander.willner@fokus.fraunhofer.de) is 
  head of the Industrial Internet of Things (IIoT) Center 
  at the Fraunhofer Institute for Open Communication Systems (FOKUS) 
  and head of the IIoT research group within the chair of Next Generation Networks (AV) 
  at the Technical University Berlin (TUB).
He is working with his groups in applying standard-based Internet of Things (IoT) technologies 
  to industrial domains. 
With a focus on moving towards the realization of interoperable communication within the Industry 4.0,
  the most important research areas include industrial real-time networks (e.g. 5G/TSN), 
  middleware systems (e.g. OPC UA), distributed AI (e.g. using Digital Twins) 
  and distributed Cloud Computing (e.g. Edge Computing) including management and orchestration.
Prior research positions include the University of Bonn, 
  Dr. Willner holds an M.Sc. and a Dr.-Ing. degree in computer science 
  from the University Göttingen and the Technical University Berlin respectively.

\sidenote{Varun Gowtham}
Varun Gowtham [M’20] (v.gowtham@tu-berlin.de) is a research fellow at the chair of
Next Generation Networks (AV), Technical University Berlin (TUB). He is
currently working in the area of real-time systems in Industry 4.0
applications using edge computing and Time-Sensitive Networking (TSN), in
particular, he is interested in software-defined distributed control systems.
He holds a dual master's degree in computer science from Technical University
Berlin, Germany and the University of Trento, Italy. Previously, he worked as
a research assistant at the Indian Institute of Science, Bengaluru.

\printbibliography%

@inproceedings{microedge2001,
author = {Gelsinger, P.P.},
booktitle = {IEEE International Solid-State Circuits Conference. Digest of Technical Papers. (ISSCC)},
doi = {10.1109/ISSCC.2001.912412},
isbn = {0-7803-6608-5},
month = {feb},
pages = {22--25},
publisher = {IEEE},
title = {{Microprocessors for the new millennium: Challenges, opportunities, and new frontiers}},
url = {http://ieeexplore.ieee.org/document/912412/},
year = {2001}
}

@article{LopezEdge2005,
author = {Lopez, Pedro Garcia and Montresor, Alberto and Epema, Dick and Datta, Anwitaman and Higashino, Teruo and Iamnitchi, Adriana and Barcellos, Marinho and Felber, Pascal and Riviere, Etienne and {Garcia Lopez}, Pedro and Montresor, Alberto and Epema, Dick and Datta, Anwitaman and Higashino, Teruo and Iamnitchi, Adriana and Barcellos, Marinho and Felber, Pascal and Riviere, Etienne},
doi = {10.1145/2831347.2831354},
issn = {0146-4833},
journal = {ACM SIGCOMM Computer Communication Review},
month = {sep},
number = {5},
pages = {37--45},
title = {{Edge-centric Computing: Vision and Challenges}},
url = {https://dl.acm.org/citation.cfm?id=2831354},
volume = {45},
year = {2005}
}

@techreport{esti-mec202wlan,
address = {Sophia Antipolis},
author = {ETSI},
institution = {European Telecommunications Standards Institute (ETSI)},
month = {jun},
title = {{GS MEC 028: Multi-access Edge Computing (MEC); WLAN Information API}},
year = {2020}
}

@techreport{iira,
author = {IIC},
institution = {Industrial Internet Consortium (IIC)},
month = {jun},
pages = {1----101},
title = {{Industrial Internet Reference Architecture}},
type = {Technical Report Version 1.9},
year = {2019}
}

@incollection{vermesan2018next,
author = {Vermesan, Ovidiu and Eisenhauer, Markus and Serrano, Martin and Guillemin, Patrick and Sundmaeker, Harald and Tragos, Elias and Valino, Javier and Copigneaux, Betrand and Presser, Mirko and Aagaard, Annabeth and Bahr, Roy and Darmois, Emmanuel},
booktitle = {Next Generation Internet of Things. Distributed Intelligence at the Edge and Human Machine-to-Machine Cooperation},
chapter = {3},
doi = {10.13052/rp-9788770220071},
editor = {Vermesan, Ovidiu and Bacquet, Jo{\"{e}}l},
isbn = {9788770220088},
month = {nov},
pages = {19 -- 102},
publisher = {River Publishers},
title = {{The Next Generation Internet of Things – Hyperconnectivity and Embedded Intelligence at the Edge}},
url = {http://riverpublishers.com/dissertations{\_}xml/9788770220071/9788770220071.xml},
year = {2018}
}

@book{zikopoulos2011understanding,
author = {Zikopoulos, Paul and Eaton, Chris and Others},
isbn = {978-0-07-179053-6},
month = {oct},
publisher = {McGraw-Hill Osborne Media},
title = {{Understanding big data: Analytics for enterprise class hadoop and streaming data}},
year = {2011}
}

@techreport{Lin2018,
author = {Lin, Shi-Wan and Murphy, Brett and Clauer, Erich and Loewen, Ulrich and Neubert, Ralf and Bachmann, Gerd and Pai, Madhusudan and Hankel, Martin},
institution = {IIC / PI4.0},
month = {dec},
pages = {19},
title = {{Architecture Alignment and Interoperability - An Industrial Internet Consortium and Plattform Industrie 4.0 Joint Whitepaper}},
url = {http://www.iiconsortium.org/pdf/JTG2{\_}Whitepaper{\_}final{\_}20171205.pdf},
year = {2018}
}

@article{bellavista2019differentiated,
author = {Bellavista, Paolo and Corradi, Antonio and Foschini, Luca and Scotece, Domenico},
doi = {10.1109/ACCESS.2019.2943848},
issn = {2169-3536},
journal = {IEEE Access},
month = {sep},
pages = {139746--139758},
publisher = {IEEE},
title = {{Differentiated Service/Data Migration for Edge Services Leveraging Container Characteristics}},
url = {https://ieeexplore.ieee.org/document/8850058/},
volume = {7},
year = {2019}
}

@article{owais2016extract,
author = {Sami, Suhail and Sael, Nada},
doi = {10.14569/IJACSA.2016.070337},
issn = {21565570},
journal = {International Journal of Advanced Computer Science and Applications},
month = {mar},
number = {3},
pages = {254--258},
title = {{Extract Five Categories CPIVW from the 9V's Characteristics of the Big Data}},
url = {http://thesai.org/Publications/ViewPaper?Volume=7{\&}Issue=3{\&}Code=ijacsa{\&}SerialNo=37},
volume = {7},
year = {2016}
}

@techreport{ngmn-5g,
author = {{NGMN Alliance}},
institution = {NGMN Alliance},
title = {{5G White Paper - Executive Version}},
year = {2014}
}

@techreport{pi4rami2015,
author = {Adolphs, Peter and Bedenbender, Heinz and Dirzus, D and Ehlich, M and Epple, U and Hankel, M and Heidel, R and Hoffmeister, M and Huhle, H and K{\"{a}}rcher, B and Others},
institution = {ZVEI / VDI / Plattform Industrie 4.0},
title = {{Reference Architecture Model Industrie 4.0 (Rami 4.0)}},
year = {2015}
}

@techreport{din91345,
author = {{Deutsches Institut f{\"{u}}r Normung}},
institution = {DIN},
month = {apr},
number = {SPEC 91345},
title = {{Referenzarchitekturmodell Industrie 4.0 (RAMI4.0)}},
year = {2016}
}

@article{edge2016,
archivePrefix = {arXiv},
arxivId = {arXiv:1011.1669v3},
author = {Shi, Weisong and Cao, Jie and Zhang, Quan and Li, Youhuizi and Xu, Lanyu},
doi = {10.1109/JIOT.2016.2579198},
eprint = {arXiv:1011.1669v3},
isbn = {2327-4662 VO - 3},
issn = {23274662},
journal = {IEEE Internet of Things Journal},
month = {oct},
number = {5},
pmid = {15003161},
title = {{Edge Computing: Vision and Challenges}},
url = {https://ieeexplore.ieee.org/abstract/document/7488250/},
volume = {3},
year = {2016}
}

@techreport{iec2017edgeintelligence,
address = {Switzerland},
author = {IEC},
howpublished = {White Paper},
institution = {International Electrotechnical Commission (IEC)},
month = {sep},
pages = {1----134},
title = {{Edge Intelligence}},
year = {2017}
}

@misc{lfedge2019,
author = {{The Linux Foundation}},
booktitle = {LF Edge},
month = {aug},
title = {{Open Glossary of Edge Computing 2.0}},
url = {https://www.stateoftheedge.com},
urldate = {2020-01-13},
year = {2019}
}

@inproceedings{willner2019edgesustainability,
address = {Potsdam},
archivePrefix = {arXiv},
arxivId = {1912.08530},
author = {Hamm, Andrea and Willner, Alexander and Schieferdecker, Ina},
booktitle = {Proceeding of WI2020},
doi = {10.30844/wi_2020_g1-hamm},
eprint = {1912.08530},
month = {mar},
pages = {694--709},
publisher = {GITO Verlag},
title = {{Edge Computing: A Comprehensive Survey of Current Initiatives and a Roadmap for a Sustainable Edge Computing Development}},
url = {http://arxiv.org/abs/1912.08530 https://library.gito.de/oa{\_}wi2020-g1.html},
year = {2020}
}

@techreport{sgam2012,
author = {{Smart Grid Coordination Group}},
institution = {CEN-CENELEC-ETSI Smart Grid Coordination Group},
month = {nov},
title = {{Smart Grid Reference Architecture}},
year = {2012}
}

\end{document}